# A new effective potential for colloidal dispersions with polymer chains emerging from mesoscopic scale interactions

**Ketzasmin A. Terrón-Mejía, Roberto López-Rendón**

Laboratorio de Bioingeniería Molecular a Multiescala, Facultad de Ciencias,
Universidad Autónoma del Estado de México,
Av. Instituto Literario 100, 50000, Toluca, Mexico

**Armando Gama Goicochea[1]**

Tecnológico de Estudios Superiores de Ecatepec, Av. Tecnológico s/n, Ecatepec, Estado de México 55210, Mexico

## ABSTRACT

A new potential of mean force is proposed for colloidal dispersions, which is obtained from coarse – grained, pair interactions between colloidal particles formed by the explicit grouping of particles that are themselves groups of atoms and molecules. Using numerical simulations, we start by constructing colloidal particles made up of 250 mesoscopic particles joined by springs and interacting with each other through short – range forces. Afterward we proceed to model several colloidal concentrations and obtain the colloidal particles pair correlation function, from which we derive the potential of mean force. In our second case study, we add linear polymer chains of the same length that attach by one of their ends to the colloids, at a fixed colloidal concentration, and carry out numerical simulations at increasing concentrations of polymer chains, from which we obtain the potential of mean force once again following the same procedure as in the previous case. For the last case study we fixed both the colloids' and the polymer chains' concentration, while increasing the length of the polymer chains and obtain again the potential of mean force. In all of these simulations, the solvent particles are included explicitly. Using these data we propose a new effective potential of interaction for colloidal dispersion whose parameters can be obtained from mesoscopic scale parameters and carry out standard molecular dynamics simulations with this new potential, thereby providing a route for more fundamentally obtained, coarse – grained approaches to model colloidal dispersions.

---

[1] Corresponding author. Electronic mail: agama@alumni.stanford.edu

# Introduction

Colloidal dispersions have numerous applications in various fields such as chemistry, biology, medicine, and engineering [1-2]. There is a need to obtain potential functions that are useful alternatives to the Lennard – Jones (LJ) and other competing potentials for the modeling of colloidal dispersions in equilibrium, so that their characteristic parameters are not based on quantum ab initio simulations of experiments on isolated molecules, but on mesoscopic scale situations, for better comparison with experiments. Interaction forces between colloidal particles in all suspensions/emulsions play an important role in determining the properties of the materials, such as the shelf life, stability, rheology and flavor, the behavior of a number of industrial processes (e.g. mixings, membrane filtrations) as well as the formula of chemical and pharmaceutical products.

Theoretical studies on the equilibrium properties of colloidal dispersions by Derjaguin and Landau [3] as well as Verwey and Overbeek [4] focused on understanding the origins of interactions between colloidal particles. This resulted in the Derjaguin–Landau–Verwey–Overbeek (DLVO) theory, which describes forces of molecular origin between bodies immersed in an electrolyte, and can be used to quantify several key features of colloid aggregation. The colloid dispersion interaction components of the DLVO model are treated by summing the pairwise dispersion interactions between the constituent atoms/molecules in the particles.

The simplest potential model for non – electrostatic interactions in colloidal particles is the hard sphere potential, in which the particles are regarded as a hard sphere are elastically reflected on contact with other particle [5]. Another simple and popular potential model is the LJ potential [6], which is useful to mimic short range repulsion and medium range van der Waals attraction. On the other hand, Hamaker [7] and de Boer [8] investigated theoretically the dispersion forces acting between colloidal objects. They considered spherical bodies, assumed pairwise additivity of interatomic dispersion energies, and demonstrated the essential results that although the range of atomic forces was of the order of atomic dimensions, the sum of the dispersion energies resulted in an interaction range for colloidal bodies of the order of their dimensions.

The classic numerical approaches used for studying colloidal dispersions, such as Monte Carlo (MC) and Molecular Dynamics (MD) [9]. However, this methods are inappropriate for describing mesoscale properties, because they still require very long simulation times and exceedingly large computer memory. To obtain a better understanding of the unique phenomenon of ordering, which occurs in colloidal suspensions, reliable mesoscopic computational models are sorely needed [10]. For systems with colloidal beds of a similar size to the complex fluid microstructures (e.g., polymeric clusters, large blood cells) or not much larger, the bed–solvent particle interactions become important. These molecular interactions are responsible for creating colloidal microstructures, such as micelles and colloidal crystals, which can be simulated within the framework of dissipative particle dynamics (DPD). The true advantage of DPD over MD consists of the possibility of matching the scale of discrete-particle simulation to the dominant spatiotemporal scales of the entire system. In MD simulation the time scales associated with evolution of heavy colloidal particles are many orders of magnitude larger than the temporal evolution of solvent particles.

Several approaches have been proposed for coupling a DPD fluid to colloidal particles. In early works based on the DPD method, colloids were simulated by constraining clusters of DPD particles to move as a rigid object. Examples of the application of DPD methods to colloidal dispersions have been reported in many sources [11-15]. In most of these works, each colloidal particle was modeled as a group of dissipative particles. However, the interactions between colloidal particles and solvent molecules in a real dispersion ought to depend on the characteristics of the dispersion of interest. In other words, such interactions are strongly dependent on the mass and diameter ratios of colloidal particles to solvent molecules, the properties of the interaction potential between such particles, etc. If we take into account that dissipative particles themselves are just virtual particles, which is a cluster or group of solvent molecules, it may be possible to use a model potential for the interaction between dissipative particles, instead of regarding a colloidal particle as a group of dissipative particles [16].

A fundamental aspect for studying colloidal dispersions under framework of DPD lies in the model interaction potential used. Although one is in principle free to choose any potential form, the conservative force is often chosen purely repulsive and linear to allow for large time-steps and hence make the DPD simulations as fast as possible. Such linear forces lead to a finite conservative potential, meaning that the centers of mass of the particles can actually overlap. This reflects the fact that effective interactions of soft objects often lead to a repulsive but finite potential [17]. An effective potential of a one-component model that accurately reproduces the colloid–colloid radial distribution function of a colloid–polymer mixture was proposed by Guzman and de Pablo [18]. Although this model was not made for studying colloidal systems mesoscopic scale, the particles of this effective model interact through an effective potential, obtained by inversion of the Ornstein–Zernike equation and a closure suited for fluids with repulsive cores. Pool and Bolhuis discussed the implication for the applicability of soft repulsive potentials for the study of micelle formation [19]. They compared results of two surfactant models: one based on LJ interactions and one based on the soft repulsive potential that is often used in dissipative particle dynamics (DPD). Meanwhile, Pan and Tartakovsky used a model where the interaction between particles is modeled by central and non-central DPD forces, which conserve both linear and angular momentums exactly [20]. Vincent et al. [21] derived an interaction potential between particles with grafted polymer chains in a solvent containing additional polymer molecules, which includes parameters such as the polymer χ-parameter, polymer adsorbed densities, molar volumes, etc. Satoh and Chantrell investigated the validity of the application of the dissipative particle dynamics (DPD) method to ferromagnetic colloidal dispersions by conducting DPD simulations for a two–dimensional system [16]. Its model interaction potential, based on LJ potential, give rise to physically reasonable aggregate structures under circumstances of strong magnetic particle–particle interactions as well as a strong external magnetic field.

Despite the efforts that have been made to model interactions colloidal mesoscopic level, the lack of efficient potential models to describing efficiently conservative forces, it is crucial to understand colloidal systems. In all the cases cited above, the LJ potential was used at the atomistic level. A good effective potential for studying colloidal dispersions must contain adjustable parameters that represent aspects such as the concentration of colloids, polymer concentration and size of the chains among others. To the date, an effective potential with such characteristics for colloidal dispersions under framework of DPD has not yet been proposed. That is the central purpose of the present work.

## Models and Methods

In MD [22], the molecules move under the action of the forces that are set up between them. These forces are pairwise, additive, and symmetric. The technique models reality on the molecular scale and could theoretically be used to accurately simulate a large macroscopic system. However, computational considerations limit the size of system and the time for which it can be simulated. From Newton's second law of motion, the evolution of the position, $\mathbf{r}_i$, and momentum, $\mathbf{p}_i = m\mathbf{v}_i$, of particle $i$ with mass, $m_i$, and velocity $\mathbf{v}$ are described by

$$\dot{\mathbf{r}}_i = \frac{\mathbf{p}_i}{m_i} \quad \text{and} \quad \dot{\mathbf{p}}_i = \mathbf{F}_i(t) = \sum_{j \neq i} \mathbf{F}_{ij}^C, \tag{1}$$

where $\mathbf{F}_i(t)$ is the force acting on particle $i$ at time $t$ and $\mathbf{F}_{ij}^C$ is the conservative force acting on particle $i$ due to particle $j$. This conservative force is simply a central force, derivable from some effective potential $\phi_{ij}$ as follows

$$\mathbf{F}_{ij}^C = -\frac{\partial \phi_{ij}}{\partial \mathbf{r}_{ij}}. \tag{2}$$

As we can see, the conservative force is only a function of particle positions, and includes inter-particle forces as well as any effects from external forces. The physical basis of conservative inter-particle interactions is typically a very short-range repulsion due to steric effects (i.e. volume exclusion/particle collision), a short-range attractive force that has its physical origins in dispersion interactions, and a long-range screened electrostatic force, which may be either attractive or repulsive. For multicomponent solvents, attractive depletion forces due to the presence of additional solutes can also be significant. Typically, surface chemistries of colloidal particles and solvent properties (pH, ionic strength) are modulated in order to balance the attractive van der Waals force with a repulsive electrostatic force and promote particle dispersion. Conservative inter-particle forces can often be approximated as a sum of pairwise two-body interactions, each of which depend only on the separation $\mathbf{r}_{ij}$ between particles $i$ and $j$ (as well as relative orientation for non-spherical particles). Determining the functional form of $\phi_{ij}$ for two particles in a solvent requires an averaging (or coarse-graining) of all degrees of freedom other than the inter-particle separation. In a rigorous statistical mechanical framework, $\phi_{ij}$ is the potential of mean force between colloidal particles [15].

The DPD method introduced by Hoogerbrugge and Koelman [23] treats a fluid using a collection of particles with 'soft' interactions, allowing for a much larger time step than atomistic MD. The characteristic dimension of DPD particles (i.e. the cutoff of the DPD interparticle interactions) is much larger than the molecular dimension of fluid molecules, but should be smaller than the relevant flow characteristic size (in the case of suspensions, the colloid particle diameter). From a molecular dynamics perspective, DPD is a traditional pairwise interaction model with a short-range cutoff and can be implemented similar to a LJ potential. The foundations of the DPD method can be found in various sources [24-28], therefore we shall only outline some general aspects of this technique. The idea behind DPD simulations is similar to a traditional molecular dynamics algorithm [22] in the sense that one must integrate Newton's second law of motion using finite time steps to obtain the particles' positions and momenta from the total force. A difference from atomistic molecular dynamics is that the DPD model involves not only a conservative force ($\mathbf{F}^C$), but also random ($\mathbf{F}^R$), and dissipative ($\mathbf{F}^D$), components acting between any two particles $i$ and $j$, placed a distance $r_{ij}$ apart. In its traditional form, the DPD total force is the sum of these three components [24], as expressed by eq. (3).

$$\dot{\mathbf{p}}_i = \mathbf{F}_i(t) = \sum_{j \neq i} \mathbf{F}_{ij}^C + \sum_{j \neq i} \mathbf{F}_{ij}^D + \sum_{j \neq i} \mathbf{F}_{ij}^R. \tag{3}$$

All forces between particles *i* and *j* are zero beyond a finite cutoff radius $r_c$, which represents the intrinsic length scale of the DPD model and is usually also chosen as $r_c \equiv 1$. The conservative force determines the thermodynamics of the DPD system and is defined by a soft repulsion:

$$\mathbf{F}_{ij}^C = \begin{cases} a_{ij}(1 - r_{ij})\hat{\mathbf{r}}_{ij} & r_{ij} \leq r_c \\ 0 & r_{ij} > r_c \end{cases}, \tag{4}$$

where $a_{ij}$ is the maximum repulsion strength between a pair of particles *i* and *j*, and **r**$_{ij}$ = **r**$_i$ − **r**$_j$, r$_{ij}$ = |**r**$_{ij}$|, $\hat{\mathbf{r}}_{ij}$ = **r**$_{ij}$/r$_{ij}$. The dissipative and the random forces are given by

$$\mathbf{F}_{ij}^D = -\gamma \omega^D(r_{ij})[\hat{\mathbf{r}}_{ij} \cdot \mathbf{v}_{ij}]\hat{\mathbf{r}}_{ij}, \tag{5}$$

and

$$\mathbf{F}_{ij}^R = \sigma \omega^R(r_{ij}) \xi_{ij} \hat{\mathbf{r}}_{ij}, \tag{6}$$

where $\gamma$ is the dissipation streng, $\sigma$ is the noise amplitude, $\omega^D$ and $\omega^R$ are distance dependent weight functions, **v**$_{ij}$ = **v**$_i$ − **v**$_j$ is the relative velocity between the particles *i* and *j*, and $\xi_{ij} = \xi_{ji}$ is a random number uniformly distributed between 0 and 1 with variance $1/\Delta t$ where $\Delta t$ is the time step of the simulation. The magnitude of the dissipative and stochastic forces are related through the fluctuation-dissipation theorem [24]:

$$\omega^D(r_{ij}) = [\omega^R(r_{ij})]^2 = max\left\{\left(1 - \frac{r_{ij}}{r_c}\right)^2, 0\right\}. \tag{7}$$

At interparticle distances larger than $r_c$, all forces are equal to zero. The strengths of the dissipative and random forces are related in a way that keeps the temperature internally fixed, $k_B T = \frac{\sigma^2}{2\gamma}$; $k_B$ being Boltzmann's constant. The natural probability distribution function of the DPD model is that of the canonical ensemble, where $N$ (the total particle number), $V$, and $T$ are kept constant. Additionally, polymers are modeled as linear chains formed by DPD beads joined by freely rotating, harmonic springs with a spring constant $K_0 = 100.0$ and an equilibrium position $r_0 = 0.7$ [29] as shown in the following equation:

$$\mathbf{F}_{ij}^{\text{spring}} = -K_0(r_{ij} - r_0)\hat{\mathbf{e}}_{ij}. \tag{8}$$

The effective potential model proposed in this work for colloidal interactions acting only between colloidal particles is a combination of the classical LJ and Morse potentials. We have named this effective potential model as KATM, in honor of first author of this work and has the following form

$$\phi_{ij}(r) = \left(\frac{\sigma}{r_{ij}}\right)^n - \alpha e^{-\beta(r_{ij}-\sigma)^2}. \tag{9}$$

This potential captures the fundamental essence of short-range interactions of LJ and Morse potentials. The first term in eq. (9) captures the short – range repulsion between colloids as in the LJ potential, while the second is responsible for the attraction between colloids that arise from elastic attraction between the polymer coatings and van der Waals interactions, as in the Morse potential. The adjustable parameters of this effective potential are $\sigma$, $n$, $\alpha$ and $\beta$, and the purpose of this work is to relate them to more basic parameters such as colloid and polymer concentrations, and the polymerization degree. In the analysis performed in here, two fundamental properties were used namely, the radial distribution function, $g(r)$, and the potential mean force (PMF), $W_{PMF}(r)$. We focus here on the latter, which is an effective pair interaction that provides important thermodynamic information about many – body systems. It can be obtained from the colloids' radial distribution functions, $g(r)$, through the relation [30]:

$$W_{PMF}(r) = -k_B T \ln[g(r)]. \tag{10}$$

**Simulation Details**

Our procedure to make colloidal particles is very simple: one starts by putting a central DPD particle at a given site and bind other particles of the same type to it, using harmonic springs, see Fig. 1. This results in a larger, composed DPD particle. The total number of individual DPD particles that make up a colloidal particle is 250. Our simulations are performed under the following conditions: the quantities $k_B T = R_C = m = 1.0$, $R_C$ is the cutoff radius and $m$ is the mass of DPD particle, all particles have the same mass and size. The density is fixed for all simulations at 3.0 DPD units and the integration time step used is $\Delta t = 0.03$. The parameters σ and γ of the DPD forces are taken as 3.0 and 4.5 respectively, and the quality of solvent elected is theta solvent with a coarse-graining degree equal to 3; this means the parameter $a_{ij} = 78.0$ DPD units for every $i$ and $j$. All simulations are carried out in the canonical ensemble. All the methodology and calculations described in this work have been performed using our simulation code, called SIMES, which is designed to study complex systems at the mesoscopic scale using graphics processors technology (GPUs) [31].

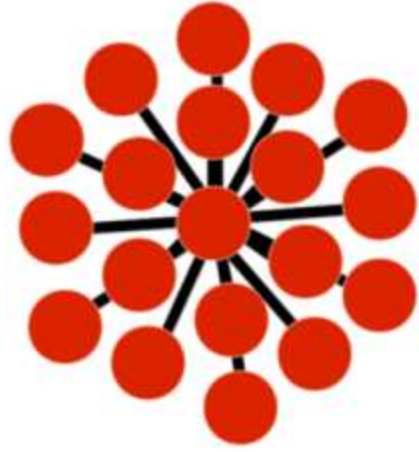

**Fig. 1** Illustration of the construction of our colloidal model particle in this work. The red particles represent DPD particles and the black lines represent the harmonic springs that bond these DPD particles to make up the colloidal particle.

In the first case of study, we explore the structure of colloidal particles as a function of their concentration. For this purpose we put 16, 32, 64, and 128 colloidal particles in a cubic simulation box whose volume was $V = 27.51^3 R_c^3$, which corresponds to colloidal concentrations equal to $\chi_{C1} = 7 \times 10^{-4}$, $\chi_{C2} = 15 \times 10^{-4}$, $\chi_{C3} = 30 \times 10^{-4}$ and $\chi_{C4} = 60 \times 10^{-4}$, respectively. The total the number of individual DPD particles making up the colloids is $N$ = 62500. For the second case study we fixed the colloidal concentration at $\chi_C = 7 \times 10^{-4}$ and we added polymer chains, which were grafted to the colloidal particles. The length of the polymer chains are fixed at $N_p$ = 6 DPD particles, bonded with harmonic springs; the constants chosen of harmonic potential are $k_0 = 100.0$ and $r_0 = 0.7 R_c$. For this case we study the impact of the polymer concentration over the structural properties of colloidal particles, the concentrations used in the simulations are $\chi_P = 0.02$, 0.04 and 0.07 in a cubic simulation cell of lateral size $L$ = 40.00 DPD units with total number of particles $N$ = 192000. The parameter $a_{c\text{-}ph}$ = 39.0 is the interaction of the polymer heads with the colloidal particle, while the parameter $a_{c\text{-}pt}$ = 78.0 controls the interaction of the polymer tail with the colloids;

both are represented by $a_{ij}$ in eq. (4). Finally, in the third case we study the impact of the length of the polymer chains on the properties of the system. For this case we fixed the concentration of colloidal particles at $\chi_C = 7 \times 10^{-4}$, and the concentration of polymer molecules at $\chi_P = 0.02$, in a cubic simulation box whose volume is $V = 40^3 R_c^3$. The polymerization degrees of the chains tested in this case are $N_p$ = 6, 12, and 24 DPD particles. We performed simulations of $10^2$ blocks of $10^4$ DPD steps for a total of $10^6$ DPD steps, reaching a total simulation time of 4.8 µs. All quantities are expressed in reduced units unless otherwise noted. In Table 1 are summarized the parameters corresponding to eq. (9) for the three case studies presented in this paper.

| Case | σ | η | α | β |
|---|---|---|---|---|
| $\chi_{c1}$ | 5.40 | 16 | 0.40 | 0.80 |
| $\chi_{c2}$ | 5.40 | 16 | 0.60 | 0.60 |
| $\chi_{c3}$ | 5.30 | 16 | 0.80 | 0.50 |
| $\chi_{p1}$ | 8.20 | 4 | 1.00 | 0.10 |
| $\chi_{p2}$ | 8.50 | 7 | 1.00 | 0.10 |
| $\chi_{p3}$ | 8.75 | 9 | 1.00 | 0.10 |
| $N_{p1}$ | 8.20 | 4 | 1.00 | 0.10 |
| $N_{p2}$ | 9.60 | 4 | 1.25 | 0.08 |
| $N_{p3}$ | 8.50 | 4 | 0.60 | 0.03 |

**Table 1**. Parameters of the KATM potential used in this work. $\chi_{cn}$, $\chi_{pn}$ and $N_{pn}$ correspond to the concentration of colloids, polymer concentration, and the size of the polymer chain respectively, subscript $n$ indicates the three case studies presented in this paper.

## Results and Discussion

The results obtained for the first case study are presented in Fig. 2; in particular, Fig. 2(a) shows the radial distribution functions of the colloidal dispersions at increasing concentrations. At all concentration except the largest one ($\chi_{C4} = 60 \times 10^{-4}$, blue line in Fig. 2(a)) the fluid behaves basically like an ideal gas with soft – core repulsion. But at the largest concentration modeled the fluid develops structure, with a large peak appearing at about $r = 5$, with periodic oscillations of about the same magnitude. The corresponding PMF in Fig. 2(b) shows the development of a shallow primary minimum at $r$ close to 5, with a secondary one appearing at $r \approx 11$. What this means is that the fluid behaves like an ideal gas until the concentration is large enough that two – body interactions become important, leading to a colloidal dispersion that can be kinetically unstable at relatively large colloidal concentrations.

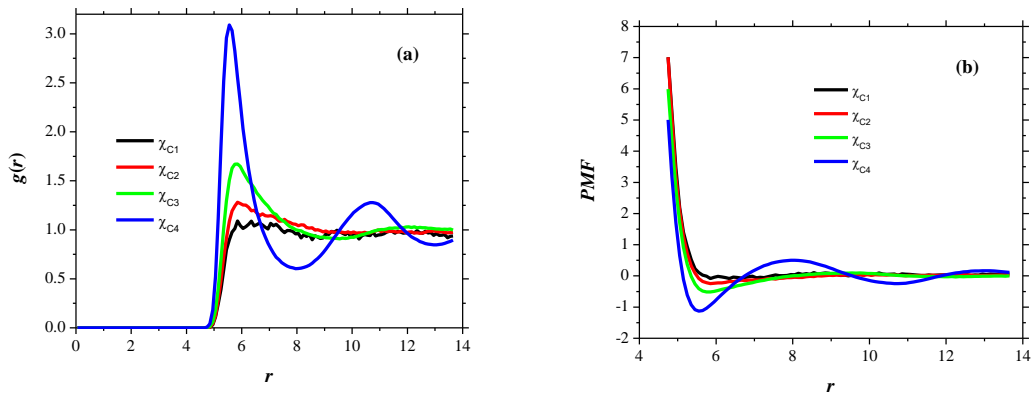

**Fig. 2** (a) Radial distribution function of the colloidal particles made up of DPD particles, at increasing concentrations of colloidal particles. (b) The corresponding potential of mean force for each colloidal concentration shown in (a); see eq. (10). See text for details. All quantities are expressed in reduced DPD units.

The snapshot of the system at the various colloidal concentration used to obtain the data shown in Fig. 2 are presented in Fig. 3.

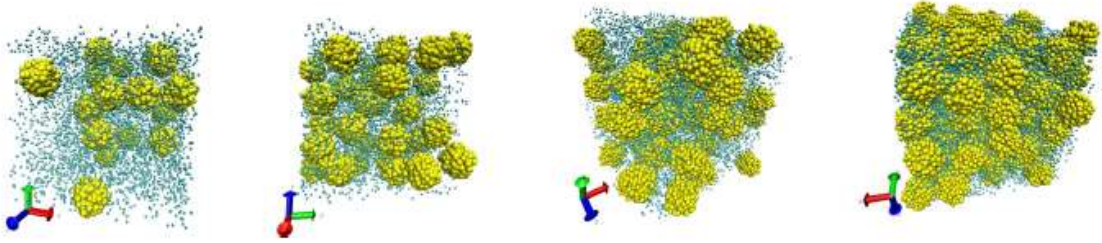

**Fig. 3** Snapshots of some typical cases of explicitly formed colloidal particles made up of DPD particles (in yellow), including the solvent monomeric particles (in cyan). These configurations correspond to colloidal concentrations $\chi_{C1}$ through $\chi_{C4}$, from left to right respectively. See text for details.

The snapshots shown in Fig, 3 show how the simulation box looks like at the four different colloidal concentration model for this fist case. It is important to notice that, even at a relatively large colloidal concentration, such as the one corresponding to the third from left snapshot in Fig. 3, there are essentially negligible many – body interactions between the composed colloidal particles. This is a consequence of the short range of the DPD interaction defined in eqs. (4) – (8).

The results of the simulations of the second case study, when the colloidal concentration was fixed and the fixed – length polymer concentration was increased are presented in Fig. 4.

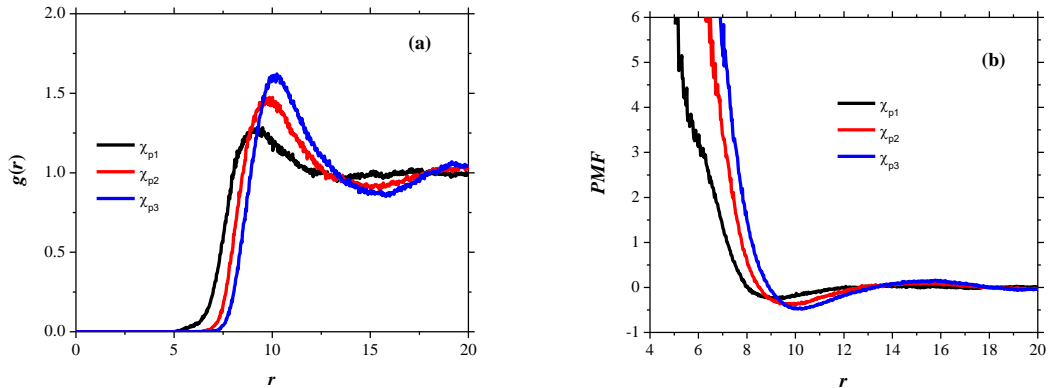

**Fig. 4** (a) Radial distribution function and (b) potential of mean force (PMF) of colloids covered by fixed – length polymer chains of increasing concentration. The concentrations of polymers are $\chi_{p1} = 0.02$ (black line), $\chi_{p2} = 0.04$ (red line), and $\chi_{p3} = 0.07$ (blue line), and the size of chains is $N_p = 6$ beads. The colloidal concentration was fixed at $\chi_C = 7 \times 10^{-4}$ (black line in Fig. 2).

It is remarkable that even for the smallest colloidal concentration, which behaves essentially as a soft – core ideal gas (see Fig. 2) there appears structuring even for the smallest polymer concentrations, as Fig.4 demonstrates. The radial distribution function (Fig, 4(a)), and hence the PMF (Fig. 4(b)) both show a weakly developing but measurable structuring for colloids at relatively short distances from each other. As Fig. 4(b) indicates, the major influence of the polymer brush coating on the colloidal particles is to increase the core repulsion between them, while at the same time developing a very shallow attractive well at relative distances close to two particle diameters ($r \approx 10$ in Fig. 4(b)). The snapshots corresponding to the simulations used to obtain the data presented in Fig. 4 are shown in Fig. 5.

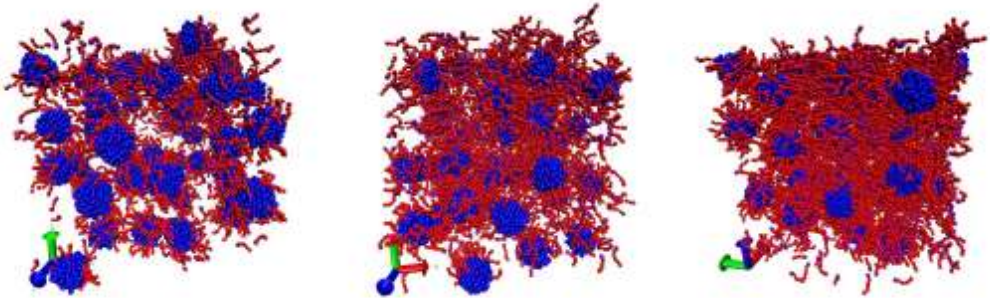

**Fig. 5** Snapshots of the simulations of DPD particle – made colloids (in blue) with grafted polymer chains (in red) of fixed length $(N_p = 6)$ at increasing polymer concentration on the colloids' surfaces. See text for details. The solvent particles are omitted for clarity.

Our last case study corresponds to the situation where colloidal and polymer concentrations are fixed, while the polymerization degree of the chains grafted to colloids is increased. The resulting radial distribution function and PMF are presented below, in Fig. 6.

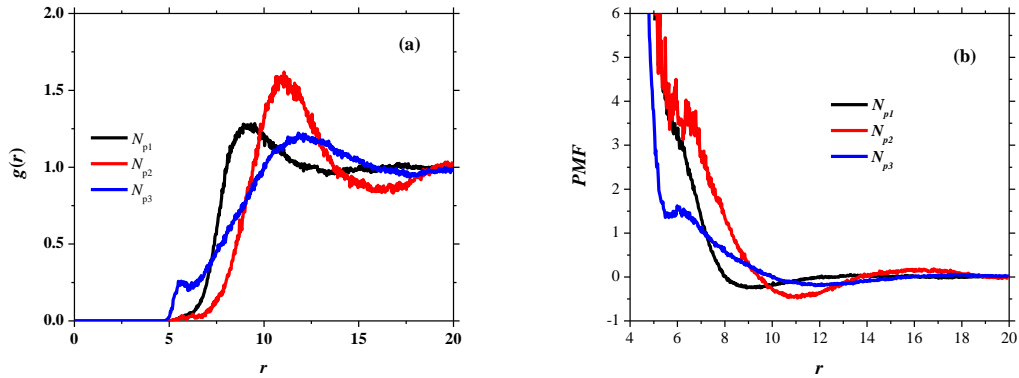

**Fig. 6** (a) Radial distribution function, and (b) PMF for a colloidal dispersion at fixed colloid and polymer concentration, while varying the polymerization degree. The polymerization degrees studied are $N_{p1} = 6$ (black line), $N_{p2} = 12$ (red line) and $N_{p3} = 24$ (blue line) beads. The colloids' concentration is fixed at $\chi_C = 7 \times 10^{-4}$, and the polymers' concentration is $\chi_P = 0.02$.

The influence of the polymerization degree on an essentially non – interacting colloidal dispersion (see black line in Fig. 2) when polymers of increasing polymerization degrees are added is quite noticeable in Fig. 6. The principal effect is the appearance of an effectively thicker colloid – particle repulsive diameter, as indicated by the red line in Fig. 6(b). Interestingly, increasing the polymerization degree up to $N_{p3} = 24$ (blue line in Fig. 6) yields to incipient colloid agglomeration, as the first maximum in the blue line in Fig. 6(a) shows. This means that a fraction of the colloids have been joined through a "bridging" mechanism, which binds particles due to the association of the long polymer chains that cover them, as is shown by the snapshot in Fig. 7; see particularly the rightmost image. When the polymerization degree leads to a radius of gyration larger than the average distance between colloids (set by the colloid concentration), the dispersant nature of the polymers is hampered and the polymers begin to act as binders, as Fig. 6(b) and Fig. 7 show.

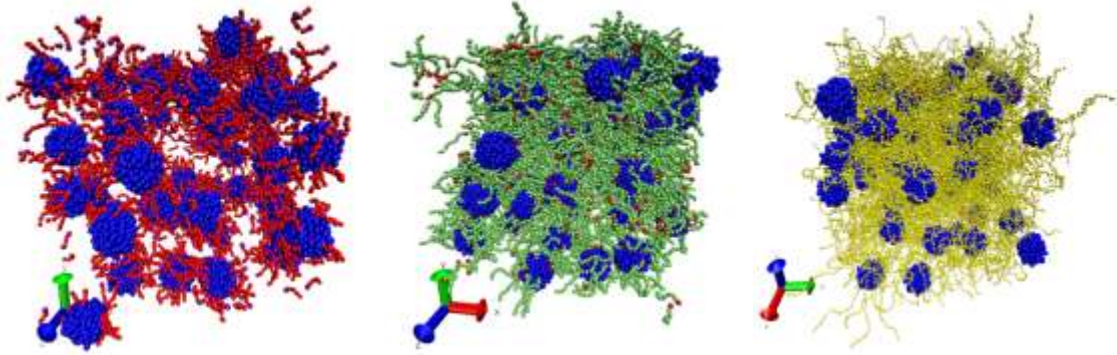

**Fig. 7** Snapshots of simulations of particle – made colloids with constant grafted density of polymer chains on the colloids' surfaces ($\chi_P = 0.02$), while increasing the polymerization degree: $N_{p1} = 6$ (red chains, leftmost snapshot), $N_{p2} = 12$ (green chains, middle snapshot) and $N_{p3} = 24$ (yellow chains, rightmost snapshot). The colloids are shown in blue.

In Fig. 8 we show the comparison between the PMF obtained from the simulations of explicitly formed colloids with polymer coatings under the various conditions described in this work, with the effective KATM potential proposed here.

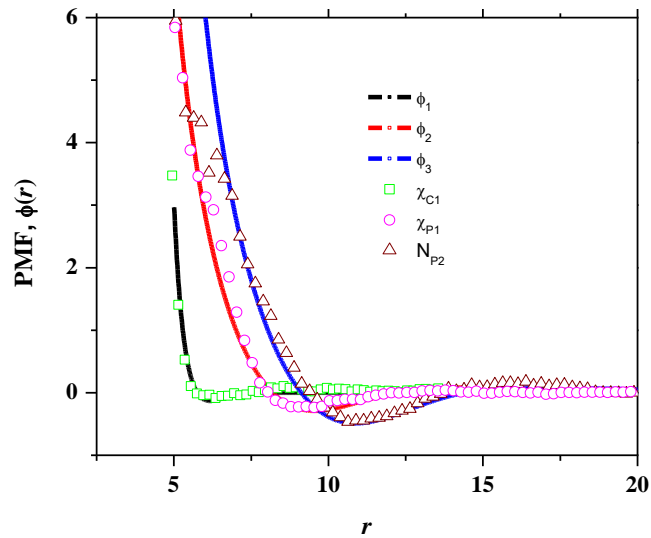

**Fig. 8** Comparison of the PMF obtained from the simulations carried out in this work for explicitly made colloids coated with polymer brushes (data points), with the effective potential ($\phi(r)$) proposed here, see eq. (9), indicated by the lines. The squares correspond to bare colloids at the concentration $\chi_{C1} = 7 \times 10^{-4}$; the circles represent the PMF for colloids at the same concentration $\chi_{C1} = 7 \times 10^{-4}$, coated with chains of constant length $N_p = 6$ beads with polymer concentration equal to $\chi_{p1} = 0.02$. The triangles represent the PMF of a colloidal dispersion whose colloidal and polymer concentrations are equal to $\chi_{C1} = 7 \times 10^{-4}$ and $\chi_{p1} = 0.02$, respectively, and the chains polymerization degree is $N_{p2} = 12$. The appropriate parameters used for the effective potential (lines) can be consulted in Table 1.

We focused on only three cases for simplicity, and as Fig. 8 shows, the match between the PMF that emerges naturally from the simulations of explicitly formed colloids with mesoscopic scale interactions is remarkably well represented by the effective potential proposed here, in eq. (9). Using the data in Table 1 we can relate the adjustable parameters of the effective potential (eq. (9)) with specific properties of the dispersion modeled, such as colloidal or polymeric concentration, as well as polymerization degree. What this means is that one can then perform ordinary, i.e. not DPD simulations between bare particles that obey the KATM potential, eq. (9), with the freedom to choose the adjustable parameters of such potential to model the colloidal concentration, the polymer concentration and the chain length by simply choosing those parameters accordingly to the trends presented here.

## Conclusions

A new effective pair potential with adjustable parameters that represent various physical situations such as colloidal concentration, polymer concentration and chain length is proposed here as an alternative for the modeling of colloidal dispersions. The novelty of the new potential is its availability to predict correctly the potential of mean force of structured colloids at the mesoscale with explicitly included polymer chains grafted on the surfaces as dispersants. The protocol presented here describes a procedure to carry out efficient simulations with bare particles that interact with the new potential proposed here without losing physical information (such as particle concentration or polymerization degree), while at the same time saving up computational resources.

## Acknowledgments

This work was supported by SIyEA-UAEM (Projects 3585/2014/CIA and 3831/2014/CIA). KATM thanks CONACyT, for a graduate student scholarship. All simulations reported in this work were performed at the Supercomputer OLINKA located at the Laboratorio de Bioingeniería Molecular a Multiescala, at the Universidad Autónoma del Estado de México. The authors are grateful to *Red de Venómica Computacional y Bioingeniería Molecular a Multiescala.* AGG would like to thank E. Blokuis (University of Leiden) for several informative conversations.